\documentclass[pra,preprint,longbibliography]{revtex4-1}
\usepackage{graphicx}	
\usepackage{appendix}
\usepackage{braket}
\usepackage{rotating}
\usepackage{simplewick}
\usepackage{amssymb}
\usepackage{amsmath}
\usepackage{txfonts}
\usepackage{bm}
\usepackage{color}
\usepackage{multirow}
\usepackage{booktabs}
\usepackage{dcolumn}

\begin{document}

\title{Numerical Evidence Invalidating Finite-Temperature Many-Body Perturbation Theory} 

\author{Punit K. Jha}
\author{So Hirata}
\email{sohirata@illinois.edu.}
\affiliation{Department of Chemistry, University of Illinois at Urbana-Champaign, Urbana, Illinois 61801, USA}

\date{\today}

\begin{abstract}
Low-order perturbation corrections to the electronic grand potential, internal energy, chemical potential, and entropy of an ideal gas of noninteracting, identical molecules at a nonzero temperature
are determined numerically as the $\lambda$-derivatives of the respective quantity 
calculated exactly (by thermal full configuration interaction) with a perturbation-scaled Hamiltonian, $\hat{H}_0 + \lambda\hat{V}$. 
The data thus obtained from the core definition of any perturbation theory serve as a benchmark against which analytical formulas can be validated.
The first- and second-order corrections from finite-temperature many-body perturbation theory discussed in many textbooks
disagree with these benchmark data. This is because the theory neglects the variation of chemical potential with $\lambda$, 
thereby failing to converge at the exact, full-interaction ($\lambda=1$) limit, unless the exact chemical potential
is known in advance. 
The renormalized finite-temperature perturbation theory [S. Hirata and X. He, J.~Chem.~Phys., {\bf 138}, 204112 (2013)] is 
also found to be incorrect.
\end{abstract}

\maketitle 

\section{Introduction}

The validity of equilibrium thermodynamics is predicated on the short-range nature of effective chemical 
interactions and the resulting randomness of the motion of constituent particles \cite{Levin}. The short-range nature, % (decaying as inverse cubic power of distance or faster)
in turn, originates from the local charge neutrality spontaneously realized in most chemical systems
and the concomitant accurate cancellation of long-range bare attractive and repulsive forces \cite{HirataARPC}. There is no reason to expect thermodynamics to work in 
a system without charge neutrality such as in a charged plasma or for a system with long-range unscreened interactions, e.g., 
a gravitational system \cite{Levin}. Note that the energy of a system with long-range interactions is not even extensive \cite{Fisher,Dyson}.

Yet, a number of modern textbooks
of quantum many-body physics \cite{bla86,thouless1972quantum,mattuck1992guide,march1995many,fetter2003quantum,negele2018quantum} dedicate an entire chapter on a finite-temperature perturbation theory for electrons that violates the charge neutrality. 
The zeroth-order (Fermi--Dirac) theory \cite{fetter2003quantum} correctly adjusts 
the zeroth-order chemical potential $\mu^{(0)}$ so as to maintain the average number of electrons at a constant value ($\bar{N}$) that exactly 
cancels the positive charge. However, at the first and higher orders, this important condition is abandoned (and the ensemble is thus altered) and 
the chemical potential is held fixed at $\mu^{(0)}$ or 
some other arbitrary value, allowing the average number of electrons to fluctuate. That this is a highly nonphysical
ensemble can be readily understood by imagining its application to a homogeneous electron gas (characterized
by a uniform electron density, $\bar{N}/V$, canceling a positive background charge) or an ideal gas of molecules 
(each of which must be electrically neutral on average). While a grand canonical ensemble with a fixed value of chemical potential may be 
useful for neutral particles, one would be hard-pressed to envision the utility of such a theory for electrons.
In fact, the finite-temperature Hartree--Fock (HF) \cite{Mermin}, density-functional \cite{PedersonJackson}, self-consistent Green's function \cite{Zgid}, 
and full-configuration-interaction (FCI) theories \cite{Kou2014} as well as the Fermi--Dirac theory \cite{fetter2003quantum} 
all adopt a grand canonical ensemble that varies the chemical potential to keep the system electrically neutral.

In this Chapter, we present benchmark data for several low-order corrections to 
the electronic grand potential, internal energy, chemical potential, and entropy of an ideal gas of identical molecules in 
a converging perturbation series that maintains the charge neutrality at any order.
They are obtained as the $\lambda$-derivatives of the respective quantities calculated exactly by the thermal FCI \cite{Kou2014}
with a perturbation-scaled Hamiltonian, $\hat{H} = \hat{H}_0 + \lambda \hat{V}$, maintaining the correct average number of electrons
at any value of $\lambda$. We show that the first and second-order perturbation corrections according to the theory given in textbooks 
disagree with these benchmark data because the theory neglects to expand the chemical potential in a perturbation series, allowing
the system to be massively charged. Consequently, the
perturbation theory described in textbooks cannot converge at the exact (FCI) limit, unless the exact chemical potential is known in advance.
The renormalized finite-temperature perturbation theory \cite{Hirata2013} is also found to be incorrect.

\section{Numerical validation}

Thermodynamic quantities such as the grand potential ($\Omega$), internal energy ($U$), chemical potential ($\mu$), and entropy ($S$) are calculated for 
a molecule unambiguously and exactly
with {\it thermal} FCI \cite{Kou2014} at any temperature ($T$) in a basis set with $m$ functions. 
First, zero-temperature FCI is executed to obtain the exact energies, $E_i^{(N,S_z)}$,
of all states of a molecule with any number of electrons ($N$) and any $S_z$ quantum number.
Second, the grand partition function, $\Xi$, is evaluated as
\begin{eqnarray}
\Xi = \sum_{N=0}^{2m} \sum_{S_z}\sum_{i} \exp\left\{-\beta \left(E_i^{(N,S_z)}-\mu N\right)\right\}
\end{eqnarray}
and
\begin{eqnarray}
\bar{N} = \frac{1}{\beta}\frac{\partial}{\partial\mu}\ln\Xi. \label{N2mu}
\end{eqnarray}
These two are the equations of state to be solved for $\Xi$ and $\mu$ simultaneously for 
a given $\beta = (k_{\text{B}}T)^{-1}$ and average number of electrons $\bar{N}$, which is 
chosen so as to keep the system electrically neutral.
Third, exact $U$ and $\Omega$ for the same $\beta$ and $\bar{N}$ are 
evaluated using
\begin{eqnarray}
\Omega &=& -\frac{1}{\beta}\ln \Xi, \\
U &=& - \frac{\partial}{\partial \beta} \ln\Xi + \mu \bar{N}.
\end{eqnarray}

The $n$th-order correction, $X^{(n)}$, of a converging perturbation series of quantity $X$ is defined \cite{fetter2003quantum,Hirata2017} as the $n$th derivative with respect to $\lambda$
of the same quantity calculated with the exact method, i.e., FCI, using a perturbation-scaled Hamiltonian, $\hat{H} = \hat{H}_0 + \lambda \hat{V}$ 
($\lambda=1$ corresponds to the fully interacting system of interest):
\begin{eqnarray}
X^{(n)} = \frac{1}{n!} \left.\frac{\partial^nX}{\partial \lambda^n}\right|_{\lambda=0}. \label{lambda}
\end{eqnarray}
Quantities $\Xi$ and $\mu$ (as well as $\Omega$, $U$, and $S$) vary with the strength of perturbation $\lambda$. 
Numerically differentiating these with respect to $\lambda$, one can obtain benchmark results for their several low-order perturbation corrections, against which analytical perturbation formulas can be 
tested and judged for their validity. 
There is a minimal risk of formulation or programming errors  in this $\lambda$-variation method \cite{Hirata2017}. One should be mindful of 
the precision of finite-difference numerical differentiations. The $\lambda$-variation method was used successfully to generate the benchmark data for the several low-order perturbation corrections 
of many-body perturbation theory, Hirschfelder--Certain degenerate perturbation theory, and Feynman--Dyson perturbation series in many-body Green's function theory \cite{Hirata2017}.

We adopt the grand canonical ensemble in line with the analytical formula \cite{fetter2003quantum,march1995many,mattuck1992guide,negele2018quantum} to be tested, although 
the canonical \cite{Kou2014} or any other ensemble can also be used. 
The method does not depend on a particular partitioning or reference wave function, either. 
Here, we employ the M{\o}ller--Plesset partitioning, where $\hat{H}_0$ is the zero-temperature Fock operator, and 
the corresponding reference is the $\bar{N}$-electron ground-state canonical HF wave function at $T=0$. Its orbitals and orbital energies are held fixed throughout the calculations.
Insofar as both the $\lambda$-variation method and analytical formulas use the identical partitioning and reference wave function,
the comparison is meaningful and it can validate or invalidate analytical formulas.
It is applied to a gas of noninteracting, identical molecules at a nonzero electronic temperature (ignoring vibrational,  rotational, and translational motions).

The thermal FCI program was verified against an independent code \cite{Kou2014} and 
several well-tested zero-temperature FCI programs. The $\lambda$-variation program used central seven-point finite-difference formulas \cite{Fornberg} at $\lambda = 0$
with a grid spacing of $\Delta\lambda=0.01$, and reproduced the M{\o}ller--Plesset perturbation energies at $T=0$. 
At each value of $\lambda$, chemical potential $\mu$ is determined by solving equation (\ref{N2mu}) by a bisection method.
As $T \to 0$, this determination becomes technically difficult, and extended-precision arithmetic was used.

\subsection{Zeroth order}
In the zeroth-order finite-temperature perturbation theory, the energy of each state is the sum of the energies ($\epsilon_p$) of orbitals occupied by electrons
plus the nuclear-repulsion energy, $E_{\text{nuc.}}$. 
The additivity of the state energies simplifies $\Omega$ and $U$ into the forms \cite{fetter2003quantum,Kou2014}:
\begin{eqnarray}
\Omega^{(0)} &=& E_{\text{nuc.}} + \frac{1}{\beta} \sum_p \ln f_p^+, \label{MP0Omega}\\
U^{(0)} &=& E_{\text{nuc.}} + \sum_p \epsilon_p f_p^-, \label{MP0U}
\end{eqnarray}
where $p$ runs over all spin-orbitals spanned by the basis set, $f_p^- =[1+\exp\{\beta(\epsilon_p-\mu^{(0)})\}]^{-1}$ is the Fermi--Dirac occupancy, and $f_p^+ = 1 - f_p^-$
is the Fermi--Dirac vacancy. Superscript `(0)' denotes a zeroth-order quantity. The chemical potential, $\mu^{(0)}$, 
is determined by solving
\begin{eqnarray}
\bar{N} = \sum_p f_p^-. \label{8}
\end{eqnarray}

\begin{table}
\caption{ \label{tab:MP0} Comparison of the analytical [equations (\ref{MP0Omega}) and (\ref{MP0U})] and numerical ($\lambda$-variation) zeroth-order grand potential ($\Omega^{(0)}$) and 
internal energy ($U^{(0)}$) as a function of temperature ($T$) for the hydrogen fluoride molecule (0.9168~\AA) in the STO-3G basis set. }
\begin{ruledtabular}
\begin{tabular}{ldddd}
&  \multicolumn{2}{c}{Analytical [equations (\ref{MP0Omega}) and (\ref{MP0U})]} & \multicolumn{2}{c}{Numerical [equation (\ref{lambda})]} \\ \cline{2-3} \cline{4-5}
$T /~\text{K}$& \multicolumn{1}{c}{$\Omega^{(0)} / E_\text{h}$} & \multicolumn{1}{c}{ $U^{(0)} / E_\text{h}$} 
& \multicolumn{1}{c}{$\Omega^{(0)} / E_\text{h}$} & \multicolumn{1}{c}{ $U^{(0)} / E_\text{h}$}  \\ \hline
$10^3$ & -53.4112 & -52.5749 & -53.4112 & -52.5749 \\
$10^4$ & -53.5117 & -52.5749 & -53.5117 & -52.5749 \\
$10^5$ & -55.6365 & -52.0166 & -55.6365 & -52.0166 \\
$10^6$ & -105.947 & -50.5964 & -105.947 & -50.5964 \\
$10^7$ & -686.703 & -45.7891 & -686.703 & -45.7891 \\
$10^8$ & -6804.94 & -42.3641 & -6804.94 & -42.3641 \\
$10^9$ & -68084.5 & -41.9453 & -68084.5 & -41.9453
\end{tabular}
\end{ruledtabular}
\end{table}

There is no question whatsoever about the validity of this Fermi--Dirac theory for a system with additive state energies.
Table \ref{tab:MP0} attests to the numerically exact agreement between the analytical (Fermi--Dirac) and numerical ($\lambda$-variation) values of $\Omega^{(0)}$ and $U^{(0)}$
at all temperatures. The agreement also underscores the validity of comparison as well as of thermodynamics applied to 
a gas of noninteracting, identical molecules.

\subsection{First order}
The first-order finite-temperature perturbation correction to $\Omega$ is given \cite{fetter2003quantum,bla86,march1995many,mattuck1992guide,negele2018quantum,SANTRA2017355,WhiteChan} as 
\begin{eqnarray}
{\Omega}^{(1)}_\text{C}=\sum_p \Delta F_{pp}(T)f_p^- -\frac{1}{2} \sum_{p,q} \braket{pq||pq} f_p^{-}f_q^{-}, 
\label{eq:F_MP1}
\end{eqnarray}
where $\braket{pq||pq}$ denotes an anti-symmetrized two-electron integral and
\begin{eqnarray}
\Delta F_{pq}(T) = H_{pq}^{\text{core}} + \sum_r \braket{pr || qr}f_r^- -\epsilon_p \delta_{pq}, \label{eq:Fock}
\end{eqnarray} 
which is a temperature shift in the thermal Fock matrix.

We place subscript `C' (standing for the conventional finite-temperature theory) to distinguish this from 
the true first-order correction ($\Omega^{(1)}$) obtainable numerically from the $\lambda$-variation method. Across textbooks and research articles, 
there seems general agreement about the right-hand-side expression, but there is some confusion over which thermodynamic property it corresponds to.
Blaizot and Ripka \cite{bla86} and Thouless \cite{thouless1972quantum}, for instance, clearly assign it to a correction to the grand potential, whereas
Mattuck \cite{mattuck1992guide} [e.g., equation (14.48) in page 249] seems to 
relate it to internal energy. 

\begin{table}
\caption{ \label{tab:HF}  Comparison of the analytical [equation (\ref{eq:F_MP1})] and numerical ($\lambda$-variation) first-order grand potential ($\Omega^{(1)}$) and 
internal energy ($U^{(1)}$) as a function of temperature ($T$) for the hydrogen fluoride molecule. The numerical (not analytical) data are the correct benchmark.}
\begin{ruledtabular}
\begin{tabular}{lddd}
& \multicolumn{1}{r}{Analytical [equation (\ref{eq:F_MP1})]} & \multicolumn{2}{c}{Numerical [equation (\ref{lambda})]} \\ \cline{2-2} \cline{3-4}
$T /~\text{K}$ &  \multicolumn{1}{c}{${\Omega}^{(1)}_\text{C} / E_\text{h}$\footnotemark[1]} & \multicolumn{1}{c}{$\Omega^{(1)} / E_\text{h}$} & \multicolumn{1}{c}{$U^{(1)} / E_\text{h}$} \\ \hline
$10^3$ & -45.9959 & -45.9959 & -45.9959 \\
$10^4$ & -45.9959 & -45.9959 & -45.9959 \\
$10^5$ & -46.0203 & -45.2684 & -45.9479 \\
$10^6$ & -46.2152 & -44.5256 & -46.1767 \\
$10^7$ & -46.1802 & -43.1991 & -46.2355 \\
$10^8$ & -46.1068 & -41.9847 & -46.1180 \\
$10^9$ & -46.0963 & -41.8264 & -46.0975 
\end{tabular}
\footnotetext[1]{Recalculated based on the formula and data reported originally by White and Chan \cite{WhiteChan}.}
\end{ruledtabular}
\end{table}

Table \ref{tab:HF} shows that ${\Omega}_\text{C}^{(1)}$ 
does not agree with either $\Omega^{(1)}$ or $U^{(1)}$ determined by the benchmark $\lambda$-variation method at high temperatures, suggesting that 
{\it the first-order formula [equation (\ref{eq:F_MP1})] is incorrect and corresponds to neither the grand potential nor the internal energy}.
This is because the conventional theory neglects 
the variation of $\mu$ with $\lambda$ (or equivalently the perturbation correction to $\mu$) that maintains 
the average number of electrons at $\bar{N}$ at all $\lambda$.

White and Chan \cite{WhiteChan}, however, showed that ${\Omega}_\text{C}^{(1)}$ agrees numerically exactly with
the $\lambda$-variation $\Omega^{(1)}$ obtained by holding $\mu$ fixed at $\mu^{(0)}$ or some other arbitrary value (such as 0~$E_\text{h}$),
which we numerically reproduced. 
Therefore, the analytical formulas of the conventional theory such as equation (\ref{eq:F_MP1}) are 
mathematically correct only under such constraints imposed on $\mu^{(0)}$.
However, its perturbation corrections are not part of a converging series towards the exact limit, i.e., 
thermal FCI with full perturbation strength ($\lambda=1$), unless the exact $\mu$ is known in advance and used 
as $\mu^{(0)}$ in the zeroth-order Hamiltonian and Fermi--Dirac function. 
This defeats the purpose of a perturbation theory 
because the exact $\mu$ can only be obtained by solving equation (\ref{N2mu}) for a given $\bar{N}$  
by the very thermal FCI procedure with $\lambda=1$. 
Nor is a perturbation theory useful if $\mu$ is chosen arbitrarily at the expense of allowing the average number of charged particles 
to fluctuate. In a macroscopic system, $\bar{N}$  needs to be held fixed at the value that maintains overall charge neutrality, without which equilibrium thermodynamics itself breaks down \cite{Levin}.
{\it Therefore, while the conventional finite-temperature perturbation theory is correct mathematically, but it is incorrect physically
as its underlying ansatz is 
unrealistic, severely curtailing its utility. Because it neglects to expand $\mu$ in a perturbation series but instead allows the average number
of charged particles to fluctuate, it fails to converge at the exact limit, applying equilibrium thermodynamics to a massively charged
macroscopic system that does not even obey its laws.}

\begin{table}
\caption{ \label{tab:mu} Perturbation corrections to the chemical potential ($\mu$) as a function of temperature ($T$) obtained by the $\lambda$-variation method for the hydrogen fluoride molecule. }
\begin{ruledtabular}

\begin{tabular}{lddd}
$T /~\text{K}$ & \multicolumn{1}{c}{$\mu^{(0)} / E_\text{h}$\footnotemark[1]} & \multicolumn{1}{c}{$\mu^{(1)} / E_\text{h}$} & \multicolumn{1}{c}{$\mu^{(2)} / E_\text{h}$} \\ \hline
$10^3$ & 0.08363   &  0.00000  & 0.04180 \\
$10^4$ & 0.09368   &  0.00000  & 0.04151 \\
$10^5$ & 0.27223   &  -0.07519 & 0.23198 \\
$10^6$ & 3.96127   &  -0.16896 & 0.08509 \\
$10^7$ & 47.1497  &  -0.29811 & 0.01775 \\
$10^8$ & 505.061 &  -0.41221 & 0.00249 \\
$10^9$ & 5092.05 & -0.42699 & 0.00026 
\end{tabular}
\footnotetext[1]{It tends to the midpoint of the highest occupied and lowest unoccupied orbital energies as $T \to 0$ \cite{Kou2014}.}
\end{ruledtabular}
\end{table}

The conventional theory is not a good approximation to the $\lambda$-variation results, either, because the neglected
$\lambda$-dependence of $\mu$ is significant, as evidenced by the nonzero values of $\mu^{(1)}$ and $\mu^{(2)}$ in Table \ref{tab:mu}.
Only at $T=0$, $\Omega^{(1)}_\text{C} = \Omega^{(1)} = U^{(1)}$ because both $\mu^{(1)}$ and $S^{(1)}$ vanish and 
\begin{eqnarray}
\Omega^{(n)} = U^{(n)} - TS^{(n)} - \mu^{(n)}\bar{N}, \label{Omega2U}
\end{eqnarray} 
for any $n$.
However, this agreement at $T=0$ does not occur unless $\mu = \mu^{(0)}$ is used in equation (\ref{eq:F_MP1}) (which is the case in Table \ref{tab:HF}); if any other value of $\mu$ is used,
$\Omega^{(1)}_\text{C}$ will have no relationship whatsoever to $\Omega^{(1)}$. 

\begin{table}
\caption{ \label{tab:S} Perturbation corrections to the entropy ($S$)  as a function of temperature ($T$) obtained by the $\lambda$-variation method for the hydrogen fluoride molecule. }
\begin{ruledtabular}

\begin{tabular}{lddd}
$T /~\text{K}$ & \multicolumn{1}{c}{$S^{(0)} / k_\text{B}$} & \multicolumn{1}{c}{$S^{(1)} / k_\text{B}$} & \multicolumn{1}{c}{$S^{(2)} / k_\text{B}$} \\ \hline
$10^3$ & 0.00000 & 0.00000 & 0.00000 \\
$10^4$ & 0.00000 & 0.00000 & 0.00000 \\
$10^5$ & 2.83441 & 0.22881 & 1.13696 \\
$10^6$ & 4.96972 & 0.01217 &-0.03361 \\
$10^7$ & 5.34979 & -0.00175&-0.00041 \\
$10^8$ & 5.40600 & -0.00004&-0.00001 \\
$10^9$ & 5.40673 & 0.00000 & 0.00000 \\
\end{tabular}
\end{ruledtabular}
\end{table}

In both low- and high-temperature limits, $S^{(n)}=0$ for $n \geq 1$ analytically \cite{Kou2014}, which has been confirmed 
numerically by the $\lambda$-variation method (Table \ref{tab:S}). This explains why $\Omega^{(1)}_\text{C}$ (evaluated with $\mu^{(0)}$) approaches $U^{(1)}$ instead of $\Omega^{(1)}$ 
at high temperatures. Since $S^{(1)} \to 0$ as $T \to \infty$ and $\Omega^{(1)}_\text{C}$ furthermore assumes (incorrectly)
$\mu^{(1)}=0$, we have also $\Omega^{(1)}_\text{C} \to U^{(1)}$ according to equation (\ref{Omega2U}). 
Generally, owing to the incorrect assumption of $\mu^{(n)}=0$ for $n \geq 1$ in the conventional theory, $\Omega^{(n)}_\text{C}$ is often deceptively 
close to $U^{(n)}$ because one of the two terms ($- \mu^{(n)}\bar{N}$) comprising the difference between $\Omega^{(n)}$ and $U^{(n)}$ in equation (\ref{Omega2U}) is missing.
This may be at least  partly responsible for the confusion of the identity of $\Omega^{(n)}_\text{C}$ in some textbooks. 

\begin{table*}
\caption{ \label{tab:GF2} Comparison of the grand potential ($\Omega$) obtained within 
various first- and second-order perturbation approximations as well as by the exact method   
as a function of temperature ($T$) for the hydrogen fluoride molecule.}
\begin{ruledtabular}
\begin{tabular}{lddddddd}
& \multicolumn{3}{c}{First order}& \multicolumn{3}{c}{Second order} & \multicolumn{1}{c}{Exact} \\ \cline{2-4} \cline{5-7} \cline{8-8}
{$T /~\text{K}$} 
&\multicolumn{1}{c}{$\sum_{i=0}^{1}{\Omega}_{\text{C}}^{(i)} / E_\text{h}$\footnotemark[1]}  
&\multicolumn{1}{c}{$\sum_{i=0}^{1}{\Omega}^{(i)} / E_\text{h}$\footnotemark[2]}  
&\multicolumn{1}{c}{$\text{HF}~/ E_\text{h}$\footnotemark[3]} 
&\multicolumn{1}{c}{$\sum_{i=0}^{2}{\Omega}_{\text{C}}^{(i)} / E_\text{h}$\footnotemark[1]}  
&\multicolumn{1}{c}{$\sum_{i=0}^{2}{\Omega}^{(i)} / E_\text{h}$\footnotemark[2]}  
&\multicolumn{1}{c}{$\text{GF2}~/ E_\text{h}$\footnotemark[3]} 
&\multicolumn{1}{c}{$\text{FCI}~/ E_\text{h}$\footnotemark[4]}  \\ \hline
$10^5$ &  -101.657 & -100.905 &  -101.021 & -101.926 & -103.486 & -103.067  & -102.107 \\
$10^6$ &  -152.162 & -150.473 &  -150.563 & -152.283 & -151.437 & -151.410  &-151.244 \\
$10^7$ &  -732.883 & -729.902 &  -729.937 & -732.905 & -730.099 & -730.100  &-730.095 \\
$10^8$ &  -6851.04 & -6846.92 &  -6846.98 & -6851.04 & -6846.95 & -6847.00  & -6847.00 
\end{tabular}
\footnotetext[1]{Recalculated based on the formula and data reported originally by White and Chan \cite{WhiteChan}.}
\footnotetext[2]{The $\lambda$-variation method (this work).}
\footnotetext[3]{Supplementary information of Welden {\it et al.}\ \cite{Zgid}. HF stands for thermal Hartree--Fock and GF2 for self-consistent second-order 
Green's function.}
\footnotetext[4]{Kou and Hirata \cite{Kou2014}.}
\end{ruledtabular}
\end{table*}

Finite-temperature HF theory may be defined by its grand potential, $\Omega^{(0)}+\Omega^{(1)}_\text{C}$, evaluated with orbitals and $\mu$ that are adjusted to make the thermal Fock matrix 
diagonal, while satisfying equation (\ref{8}). Owing to the latter provision that guarantees the correct average number of electrons, 
the finite-temperature HF grand potentials \cite{Zgid}, shown in Table \ref{tab:GF2}, are numerically closer to 
$\Omega^{(0)}+\Omega^{(1)}$ of the $\lambda$-variation method than to $\Omega^{(0)}+\Omega^{(1)}_\text{C}$ evaluated with a fixed $\mu^{(0)}$. 
The HF  and $\lambda$-variation results are not identical because the former is not a perturbation theory and determines $\mu$ differently. 
Nevertheless, the mutual consistency in the numerical results between the finite-temperature HF and $\lambda$-variation methods  
supports the ansatz that varies $\mu$ to maintain $\bar{N}$, but not vice versa.

\subsection{Second order}

The second-order perturbation correction to the grand potential at a nonzero temperature \cite{Kohn1960,bla86,fetter2003quantum,negele2018quantum,SANTRA2017355,WhiteChan} is given by
\begin{eqnarray}
{\Omega}_{\text{C}}^{(2)} &=&\frac{1}{4} \sum_{D \neq 0} \frac{ |\braket{pq||rs}|^2 f_p^+ f_q^+ f_r^-f_s^-} {\epsilon_r+\epsilon_s-\epsilon_p - \epsilon_q} 
\nonumber\\ && 
- \frac{\beta}{8} \sum_{D = 0} |\braket{pq||rs}|^2 f_p^+ f_q^+ f_r^-f_s^-
%\nonumber\\&& 
+ \sum_{D \neq 0} \frac{| \Delta F_{pq} (T)|^2  f_p^+ f_q^-}{\epsilon_q - \epsilon_p}
\nonumber\\&&
- \frac{\beta}{2} \sum_{D=0} | \Delta  F_{pq} (T)|^2  f_p^+  f_q^-,
 \label{eq:A2}
\end{eqnarray}
where $D \neq 0$ and $D=0$ indicate that the summation is limited
to the summands of which the denominator of the parent term 
($\epsilon_r+\epsilon_s-\epsilon_p - \epsilon_q$ or $\epsilon_q - \epsilon_p$) is nonzero and zero,
respectively. 

\begin{table}
\caption{ \label{tab:MP2} Comparison of the second-order grand potentials ($\Omega^{(2)}$) and 
internal energies ($U^{(2)}$) obtained by various analytical formulas or numerically by the $\lambda$-variation method 
as a function of temperature ($T$) for the hydrogen fluoride molecule. The numerical (not analytical) data are the correct benchmark.}
\begin{ruledtabular}
\begin{tabular}{ldddd}
& \multicolumn{2}{c}{Analytical [equations (\ref{eq:A2}) and (\ref{eq:R2})]} & \multicolumn{2}{c}{Numerical [equation (\ref{lambda})]} \\ \cline{2-3} \cline{4-5}
{$T /~\text{K}$} 
& \multicolumn{1}{c}{${\Omega}_{\text{C}}^{(2)} / E_\text{h}$\footnotemark[1]} 
& \multicolumn{1}{c}{${U}_{\text{R}}^{(2)} / E_\text{h}$} 
& \multicolumn{1}{c}{${\Omega}^{(2)} / E_\text{h}$} 
& \multicolumn{1}{c}{$U^{(2)} / E_\text{h}$}  \\ \hline
$10^3$  &  -0.01734  &-0.01734  & -0.43534 & -0.01734 \\
$10^4$  &  -0.01734  &-0.01734  & -0.43244 & -0.01734 \\
$10^5$  &  -0.26894  &-0.24287  & -2.58146 &  0.09842 \\
$10^6$  &  -0.12056  & 3.06683  & -0.96432 & -0.21984 \\
$10^7$  &  -0.02184  & 1.77859  & -0.19697 & -0.03260 \\
$10^8$  &  -0.00318  & 1.01395  & -0.02759 & -0.00536 \\
$10^9$  &  -0.00033  & 0.94969  & -0.00285 & -0.00057
\end{tabular}
\footnotetext[1]{Recalculated based on the formula and data reported originally by White and Chan \cite{WhiteChan}.}
\end{ruledtabular}
\end{table}

One of the present authors with a coauthor proposed \cite{Hirata2013} another expression of second-order correction to the internal energy,
\begin{eqnarray}
{U}_{\text{R}}^{(2)}&=&\frac{1}{4} \sum_{p,q,r,s} \frac{ |\braket{pq||rs}|^2 f_p^+ f_q^+ f_r^-f_s^-} {f_r^- \epsilon_r+ f_s^- \epsilon_s-f_p^+ \epsilon_p -f_q^+ \epsilon_q}
\nonumber\\
&& +\sum_{p,q} \frac{| \Delta F_{pq}(T) |^2  f_p^+ f_q^-}{f_q^-\epsilon_q - f_p^+\epsilon_p},
\label{eq:R2}
\end{eqnarray}
which differ from the conventional formula in that the temperature effect is applied symmetrically on the interactions in the numerators and denominators. 
Divergent summands generally do not occur. 
Subscript `R' stands for the renormalized finite-temperature perturbation theory \cite{Hirata2013}. 

Table \ref{tab:MP2} suggests that 
{\it both second-order correction formulas tested here are incorrect; neither is part of a converging perturbation series towards the exact (thermal FCI) limit}. 

To be specific, ${\Omega}_{\text{C}}^{(2)}$ does not agree with the benchmark $\Omega^{(2)}$ values from the $\lambda$-variation method at any temperature, but 
instead tends to agree with $U^{(2)}$ at low temperatures (which is expected from its mathematical form).
Our alternative formula, ${U}_{\text{R}}^{(2)}$, does not match $U^{(2)}$ except at low temperatures 
and is completely different from $\Omega^{(2)}$ at any temperatures studied. Therefore, the renormalized 
finite-temperature perturbation theory \cite{Hirata2013} is clearly incorrect. 

In ref.\ \cite{WhiteChan}, White and Chan again showed that ${\Omega}_{\text{C}}^{(2)}$ agrees numerically with the $\lambda$-variation $\Omega^{(2)}$ when $\mu$ 
is held fixed at a constant (such as $\mu^{(0)}$ in their calculations), which we also reproduced.
Therefore, the conventional second-order formula, ${\Omega}_{\text{C}}^{(2)}$, is mathematically correct, but only for an unrealistic and oversimplified ansatz, which
imposes $\mu^{(n)} = 0$ for $n \geq 1$. 
It is not part of a perturbation series converging at the exact ($\lambda=1$) limit, where $\mu$ is no longer $\mu^{(0)}$. 

Table \ref{tab:mu} shows that even at the lowest temperature tested ($T=10^3$~K), $\mu^{(2)}$ has a substantial value of  $0.041801~E_{\text{h}}$, 
which explains the rather large difference between $\Omega^{(2)}$ and $U^{(2)}$, which are related to each other by equation (\ref{Omega2U}) (note $S^{(2)} = 0$ and $\bar{N}=10$). 
As a result of the assumption of $\mu^{(2)}=0$, the conventional formula (${\Omega}_{\text{C}}^{(2)}$) and the $\lambda$-variation result of 
White and Chan at $T=10^3$~K are far from the true $\Omega^{(2)}$ and are closer to 
$U^{(2)}$ (or to some arbitrary value if the fixed value of $\mu$ is chosen arbitrarily). 

We also note that the sum of our benchmark $\Omega^{(n)}$ over $n=0, 1, 2, 3$ gives a close approximation to the exact 
$\Omega$ obtained by thermal FCI ($\lambda=1$) at each temperature (not shown; see Table \ref{tab:GF2} for the sums over $n=0,1,2$). This is also the case with
other quantities such as $U$, $\mu$, and $S$.
Such convergence cannot be expected from the conventional theory or White and Chan's $\lambda$-variation calculation,
unless the exact $\mu$ is known in advance and used as $\mu^{(0)}$ in their zeroth-order Hamiltonian and Fermi--Dirac function. It is rather doubtful if such a perturbation theory has much utility.

The self-consistent Green's function theory of Welden {\it et al.}\ \cite{Zgid} makes a second-order correction 
to the grand potential at a nonzero temperature in such a way that the correct average number of electrons is maintained by adjusting $\mu$. 
Their data, reproduced in Table \ref{tab:GF2}, are closer to $\Omega^{(0)}+\Omega^{(1)}+\Omega^{(2)}$ obtained from 
the $\lambda$-variation method than the results from the conventional theory. 
However, they are not identical, indicating that the theory of Welden {\it et al.}\ forms another potentially converging series that
differs from the canonical perturbation series defined by equation (\ref{lambda}). Nonetheless, the overall numerical consistency seen among the $\lambda$-variation method,
finite-temperature HF theory, self-consistent Green's function theory, and thermal FCI underscores the soundness of the ansatz that varies $\mu$ to keep $\bar{N}$ constant.
The conventional theory given in textbooks, which varies $\bar{N}$ for a fixed $\mu$, is a prominent outlier.

\section{Conclusion}

Corrections to the grand potential or internal energy calculated with
the finite-temperature first- or second-order perturbation theories proposed so far disagree with the benchmark $\lambda$-variation results. 
Exact numerical agreement at the zeroth order justifies the comparison itself. 
The disagreement stems from the fact that 
the conventional theory fails to account for a continuous change in $\mu$ as the perturbation strength $\lambda$
is raised to unity, whereupon the system becomes the true interacting system having the correct average number of electrons.
The conventional theory does not converge at this exact limit unless the exact $\mu$ is known in advance. 
Its numerical data (obtained with $\mu^{(0)}$) are an outlier in the dataset from thermal FCI, $\lambda$-variation, finite-temperature HF, and 
self-consistent Green's function theories, which are mutually consistent (but not the same) with one another by virtue of considering the variation of
$\mu$ to keep $\bar{N}$ constant at all temperatures and perturbation strengths.
While the conventional theory is mathematically correct and may be argued to be useful in some limited circumstances, 
a correct finite-temperature perturbation theory \cite{HirataJha} that 
also expands $\mu$ in a converging series should be developed for more realistic physics. 
The benchmark data presented here and the computational machinery \cite{Hirata2017} to generate them should be valuable for such endeavor.

\acknowledgments
This work was supported by the Center for Scalable, Predictive methods for Excitation and Correlated phenomena (SPEC), which is funded by 
the U.S. Department of Energy, Office of Science, Office of Basic Energy Sciences, Chemical Sciences, Geosciences, and Biosciences Division, as a part of the Computational Chemical Sciences Program
and also by the U.S. Department of Energy, Office of Science, Office of Basic Energy Sciences under Grant No.\ DE-SC0006028.
We sincerely thank Dr.\ Alec F. White and Dr.\ Garnet K.-L. Chan for many corrections to earlier drafts of this paper via ref.\ \cite{WhiteChan}.
We also thank Mr.\ Alexander Doran, Dr.\ Alexander Kunitsa, Dr.\ Debashis Mukherjee, Dr.\ Mark Pederson,
Dr.\ Robin Santra for helpful discussions.

%\bibliography{ARCC.bib}
\end{document}